\documentclass[prd,twocolumn,amsmath,amssymb,aps]{revtex4-2}
 
\usepackage{hyperref}

\newcommand{\cN}{{\cal N}}

\begin{document}

\title{Analytic action principle in $\cN=2$ AdS$_4$ harmonic superspace}

\author{Timothy Gargett}
\email{timothy.gargett@research.uwa.edu.au}

\author{Igor Samsonov}
\altaffiliation{School of Physics, University of New South Wales, Sydney 2052, Australia (present address, effective from 6 October 2025)}
\email{igor.samsonov@unsw.edu.au}

\affiliation{Department of Physics, The University of Western Australia, 35 Stirling Highway, Perth W.A. 6009, Australia}

\begin{abstract}
We develop an analytic action principle in $\mathcal N=2$ AdS$_4$ harmonic superspace and apply it to study the component structure of the free $q$-hypermultiplet model.
\end{abstract}

\maketitle


\section{Introduction and conclusions}

Superspace formulations of supersymmetric field theories provide explicit off-shell realizations of the Poincaré superalgebra, whose closure does not require the equations of motion. Such formulations are particularly advantageous for models with extended supersymmetry, since hypermultiplet matter theories require an infinite number of auxiliary fields off shell, which naturally emerge within the harmonic \cite{Galperin:1984av,HSS} and projective \cite{Gonzalez-Rey:1997pxs,Gonzalez-Rey:1997msl} superspace approaches. Being complementary \cite{Kuzenko:1998xm}, these two formalisms are especially useful for constructing classical actions of higher-spin field theories with extended supersymmetry \cite{Kuzenko:2021pqm,Kuzenko:2023vgf,Hutchings:2023iza,Kuzenko:2024vms,Buchbinder:2021ite,Buchbinder:2022kzl,Buchbinder:2022vra,Buchbinder:2024pjm,Ivanov:2024bsb,Buchbinder:2025yef}. Generalizing $\cN=2$ supersymmetric field theory to the 4D anti de-Sitter background (AdS$_4$), however, represents a nontrivial problem, as the corresponding geometry in the harmonic superspace has not yet been fully developed. Although general supergravity theories were formulated in 4D $\mathcal{N}=2$ harmonic superspace (see Refs.~\cite{Galperin:1984av,Galperin:1987em,Galperin:1987ek,Delamotte:1987yn,Zupnik:1998td,Butter:2015nza,Ivanov:2024gjo}), the specific features of $\cN=2$ AdS$_4$ harmonic superspace remain insufficiently explored, and the construction of classical actions for gauge and matter theories in these superspaces has not yet been properly addressed. In contrast, the analytic action principle in the AdS$_5$ harmonic superspace was developed in the works \cite{Kuzenko:2007aj,Kuzenko:2007vs}.

An important step toward formulating field theories in $\cN=2$ AdS$_4$ harmonic superspace was taken in Ref.~\cite{Ivanov:2025jdp}, where the analytic superspace measure was explicitly constructed. This measure plays a central role in building superfield actions for matter hypermultiplets and their interaction vertices with gauge superfields. The approach of Ref.~\cite{Ivanov:2025jdp}, however, is coordinate specific and provides limited insight into the underlying AdS$^{4|8}$ superspace geometry. In the present paper, we extend the construction of Ref.~\cite{Ivanov:2025jdp} by developing an analytic action principle in $\cN=2$ AdS$_4$ harmonic superspace in a coordinate-independent and manifestly covariant form. As an application, we derive the component structure of the free $q$-hypermultiplet model in $\cN=2$ AdS$_4$ harmonic superspace.

Despite the progress achieved in formulating field theories in $\cN=2$ AdS$_4$ harmonic superspace in the present paper and in Ref.~\cite{Ivanov:2025jdp}, the problem remains far from complete. The approach of Ref.~\cite{Ivanov:2025jdp} is based on manifestly analytic superfields and employs the so-called analytic basis in superspace. In the present work, we use the central basis, in which analyticity is not manifest but is instead imposed through superfield constraints. This latter formulation is, however, fully covariant and explicitly geometric, owing to the developments in Refs.~\cite{Kuzenko:2008qw,Kuzenko:2008ep,Kuzenko:2008ry,Butter:2011ym,Butter:2011kf,Butter:2012jj,Koning:2023ruq,Koning:2024iiq} where AdS$^{4|8}$ superspace was obtained as a particular solution of $\mathcal{N}=2$ superfield supergravity. In general $\mathcal{N}=2$ supergravity, these two coordinate bases are related by the so-called bridge superfield, originally constructed in Ref.~\cite{Butter:2015nza}. The explicit construction of this bridge superfield for the $\cN=2$ AdS$_4$ harmonic superspace background remains an important open problem, which will be addressed elsewhere.

The rest of the paper is organized as follows. In the next section, we present some background information about the geometry of the AdS$^{4|8}$ superspace and its isometry generated by the Killing vectors. In Sec.~\ref{SecHarmonic} we extend this superspace with harmonic variables and introduce analytic superspace projectors. In Sec.~\ref{SecAction} we show that these projectors may be used for construction a superfield action for an analytic Lagrangian in the harmonic superspace. We develop also the analytic action principle in a fully covariant and coordinate independent form. Then we apply this action principle for calculating the component structure of the classical action in the free $q$-hypermultiplet model. In our work, we follow superspace conventions adopted in Refs.~\cite{Kuzenko:2008qw,Kuzenko:2008ep,Kuzenko:2008ry}.

\section{Isometries of \texorpdfstring{AdS$^{4|8}$}{AdS4} superspace}

Consider a curved 4D $\cN=2$ superspace parametrized by local coordinates $z^M = (x^m,\theta^\mu_i,\bar\theta^i_{\dot\mu})$, where $m=0,1,2,3$, $\mu=1,2$, $\dot\mu=1,2$ and $i=1,2$. The Grassmann variables are related to each other by the complex conjugation, $\overline{\theta_i^\mu} = \bar\theta^{\dot\mu i}$. The supergravity structure group is chosen to be $SO(3,1)\times SU(2)$, and the covariant derivatives have the form
\begin{equation}
  {\cal D}_A = E_A + \frac12\Omega_A{}^{bc}M_{bc} + \Phi_A{}^{kl}J_{kl}\,,
\end{equation}
where $E_A = E_A{}^M\partial_M$ is the supervielbein and $M_{ab}=-M_{ba}$ are Lorentz group generators with commutation relations
\begin{equation}
  [M_{ab},M_{cd}] = 2\eta_{c[a}M_{b]d} - 2\eta_{d[a}M_{b]c}\,.
\end{equation}
They act on vectors and spinors by the rule
\begin{subequations}
\begin{align}
  [M_{ab} , {\cal D}_c] = &2\eta_{c[a}{\cal D}_{b]}\,,\\
  [M_{\alpha\beta},{\cal D}^i_{\gamma}] = &\varepsilon_{\gamma(\alpha}{\cal D}^i_{\beta)}\,,\\
  [\bar M_{\dot\alpha\dot\beta}, \bar{\cal D}^i_{\dot\gamma}] =& \varepsilon_{\dot\gamma(\dot\alpha}\bar{\cal D}^i_{\dot\beta)}\,.
\end{align}
\end{subequations}
We use the following conventions for the relations between the Lorentz generators with vector and spinor indices:
\begin{subequations}
\begin{align}
M_{ab} = &(\sigma_{ab})^{\alpha\beta}M_{\alpha\beta} - (\tilde\sigma_{ab})^{\dot\alpha\dot\beta}\bar M_{\dot\alpha\dot\beta}\,,\\
M_{\alpha\beta} = &\frac12(\sigma^{ab})_{\alpha\beta}M_{ab}\,,\\
\bar M_{\dot\alpha\dot\beta} =& -\frac12(\tilde\sigma^{ab})_{\dot\alpha\dot\beta} M_{ab}\,.
\end{align}
\end{subequations}

The $SU(2)$ generators obey
\begin{equation}
  [J^i{}_j,J^k{}_l] = \delta^i_lJ^k{}_j - \delta^k_j J^i{}_l\,,
\end{equation}
and
\begin{equation}
  [J_{kl}, {\cal D}^i_\alpha] = -\delta^i_{(k}{\cal D}_{\alpha l)}\,,\qquad
  [J_{kl}, \bar{\cal D}_i^{\dot\alpha}] = -\varepsilon_{i(k}\bar{\cal D}^{\dot\alpha}_{l)}\,.
\end{equation}

AdS$^{4|8}$ superspace may be considered as a particular solution of the $\cN=2$ supergravity with covariant derivatives obeying the following graded commutation relations
\begin{subequations}
  \label{D-algebra}
\begin{align}
  \{ {\cal D}^i_\alpha , {\cal D}^j_\beta \} =& 4 S^{ij}M_{\alpha\beta} + 2\varepsilon_{\alpha\beta}\varepsilon^{ij} S^{kl} J_{kl}\,,\label{DD}\\
  \{ \bar{\cal D}_{\dot\alpha i}, \bar{\cal D}_{\dot\beta j} \} =& - 4 S_{ij}\bar M_{\dot\alpha\dot\beta} + 2\varepsilon_{\dot\alpha\dot\beta}\varepsilon_{ij} S^{kl} J_{kl}\,,\label{barDbarD}\\
  \{ {\cal D}^i_\alpha , \bar{\cal D}_j^{\dot\beta}\} =& -2i\delta^i_j(\sigma^c)_\alpha{}^{\dot\beta}{\cal D}_c\,,\\
  [{\cal D}_a , {\cal D}^j_\beta] =& \frac i2(\sigma_a)_{\beta\dot\gamma}S^{jk}\bar{\cal D}^{\dot\gamma}_k\,,\\ 
   [{\cal D}_a, \bar{\cal D}_j^{\dot\beta}] =& \frac i2(\sigma_a)^{\alpha\dot\beta} S_{jk}{\cal D}^k_{\alpha}\,,\\
  [{\cal D}_a,{\cal D}_b] =& -S^2 M_{ab}\,,
  \label{curvature}
\end{align}
\end{subequations}
where $S^2 = \frac12S^{kl}S_{kl}$, and the $SU(2)$ connection obeys the constraints
\begin{equation}
    {\cal D}_A S^{ij} = 0\,,\qquad
    \overline{S^{ij}} = S_{ij}\,.
    \label{Sconstraints}
\end{equation}
In Ref.~\cite{Kuzenko:2008qw}, it was shown that one can impose a special gauge on the $SU(2)$ connection such that the torsion $S^{ij}$ becomes a constant, $S^{ij}= c^{ij} =\mathrm{const}$. These constants $c^{ij}$ play a central role in the construction of the AdS$_4$ harmonic superspace in Ref.~\cite{Ivanov:2025jdp}. In the present work, we do not impose this gauge and therefore keep the torsion $S^{ij}$ covariantly constant.

The isometry group of AdS$^{4|8}$ is $OSp(2|4)$. The infinitesimal isometry transformations are generated by the Killing supervector $\xi^A = (\xi^a , \xi_i^\alpha,\bar\xi^i_{\dot\alpha})$ such that the operator
\begin{equation}
  \xi = \xi^A {\cal D}_A = \xi^a {\cal D}_a + \xi^\alpha_i{\cal D}^i_\alpha + \bar\xi^i_{\dot\alpha}\bar{\cal D}_i^{\dot\alpha} 
\end{equation}
obeys
\begin{equation}
  \left[ \delta_\xi,{\cal D}_A \right] = \left[ \xi + \frac12\lambda^{ab}M_{ab} + \lambda^{ij}J_{ij} ,{\cal D}_A \right] = 0\,,
\label{KillingVectorEquation}
\end{equation}
for some real antisymmetric tensor $\lambda^{cd}(z)=-\lambda^{dc}(z)$ and real symmetric tensor $\lambda^{ij} = \overline{\lambda_{ij}}$. 

In Ref.~\cite{Kuzenko:2008qw}, it was shown that the parameters $\lambda^{ij}$ may be expressed via a single superfield $\rho$. Indeed, $S^{ij}$ being a component of supertorsion should be invariant under the isometry transformation, $\left[ \delta_\xi,S^{ij} \right]=0$. Since in the AdS$^{4|8}$ superspace this supertorsion is covariantly constant, see Eq.~(\ref{Sconstraints}), we find that $S^{ij}$ is invariant under the $SU(2)$ action of $\lambda^{ij}J_{ij}$, which implies $\lambda^{ij} =\rho S^{ij}$. This superfield, as well as other components of the Killing vector $\xi^A$ obey a number of constraints which follow from Eq.~(\ref{KillingVectorEquation}):
\begin{subequations}
\begin{align}
{\cal D}^i_\alpha \xi^\beta_j - \rho S^i{}_j \delta_\alpha^\beta - \lambda_\alpha{}^\beta\delta^i_j = &0\,,\\
\bar{\cal D}_i^{\dot\alpha} \xi^\beta_j - \frac i2 S_{ij}\xi^{\beta\dot\alpha} =&0\,,\\
\bar{\cal D}_i^{\dot\alpha} \xi^b + 2i(\sigma^b)_\beta{}^{\dot\alpha}\xi_i^\beta =&0\,,\label{xi-equation}\\
{\cal D}^i_\alpha \lambda^{cd} - 4S^{ij}\xi_j^\beta(\sigma^{cd})_{\alpha\beta} =&0\,,\\
{\cal D}^i_\alpha \rho - 2\xi^i_\alpha =&0\,.
\end{align}
\end{subequations}
These equations imply that all the components of the Killing vector and $\lambda$-parameters are expressed via a single real scalar superfield $\rho$ \cite{Kuzenko:2008qw}:
\begin{align}
  \xi_i^\alpha =& \frac12{\cal D}_i^\alpha \rho\,,& \xi_{\alpha\dot\beta} =& \frac{i}{2S^2}S_{ij}{\cal D}^i_\alpha \bar{\cal D}^j_{\dot\beta}\rho\,,\\
  \lambda^{ij} = & \rho S^{ij}\,,& 
  \lambda_{\alpha\beta} =& \frac14 {\cal D}^k_\alpha {\cal D}_{\beta k}\rho\,.
\end{align}
The superfield $\rho$ obeys the following constraints:
\begin{subequations}
\begin{align}
\left( {\cal D}^{\alpha i}{\cal D}^j_{\alpha} + 4 S^{ij} \right)\rho= &0\,,\\
\left( {\cal D}^i_\alpha\bar{\cal D}^j_{\dot\beta} - \frac1{2S^2}S^{ij}S_{kl}{\cal D}^k_\alpha\bar{\cal D}^l_{\dot\beta} \right)\rho =&0\,,\\
{\cal D}_a \rho=&0\,.
\end{align}
\end{subequations}

By definition, a superfield $\Phi$ (with Lorentz and $SU(2)$ indices suppressed) on AdS$^{4|8}$ transforms under the isometry by the rule 
\begin{equation}
\delta_\xi \Phi = \left( \xi + \frac12\lambda^{ab}M_{ab} + \rho S^{ij}J_{ij}\right) \Phi\,.
\label{delta-H}
\end{equation}
Below we will generalize these transformations to the superfields in the harmonic superspace.

\section{Introducing harmonics}
\label{SecHarmonic}

Consider $SU(2)$ harmonic variables with defining properties $u^{+i}u^-_j - u^{-i}u^+_j = \delta^i_j$, $u^{+i}u^+_i = u^{-i}u^-_i=0$. Associated with the harmonic coordinates are the covariant harmonic derivatives $D^{++}$, $D^{--}$ and $D^0$ with commutation relations of the $su(2)$ Lie algebra
\begin{subequations}
\begin{align}
  [D^{++}, D^{--}] = &D^0\,,\\
  [D^0, D^{++}] = &2D^{++}\,,\\
  [D^0, D^{--}] = &-2D^{--}\,.
\label{harmonic-derivatives-algebra}
\end{align}
\end{subequations}

Following the routine of the harmonic superspace approach \cite{Galperin:1984av}, with every iso-tensor superfield with no Lorentz indices 
\begin{equation}
\Phi^{i_1\ldots i_{m+n}}(z) = \Phi^{(i_1\ldots i_{m+n})}(z)\,,
\label{Phi-isotensor}
\end{equation}
we associate a harmonic superfield
\begin{equation}
	\Phi^{(m,n)}(z,u) = u^+_{i_1} \ldots u^+_{i_m} u^-_{i_{m+1}} \ldots u^-_{i_{m+n}} \Phi^{i_1\ldots i_{m+n}}(z)\,.
\end{equation}
By construction this superfield has $U(1)$ charge $q=m-n$,
\begin{equation}
    D^0\Phi^{(m,n)} = (m-n)\Phi^{(m,n)}\,,
\end{equation}
and, thus, it is a smooth function over the homogeneous space $S^2\simeq SU(2)/U(1)$. More generally, given a set of superfields with arbitrary number of $SU(2)$ indices of the type (\ref{Phi-isotensor}), we define a general harmonic superfield with a given $U(1)$ charge $q$:
\begin{equation}
	\Phi^{(q)}(z,u) = 
	\begin{cases}
		\sum_{n\ge0} \Phi^{(n+q,n)}(z,u) & q\ge 0\\
		\sum_{n\ge0} \Phi^{(n,n-q)}(z,u) & q< 0\,,
	\end{cases}
	\label{Phi-q}
\end{equation}
\begin{equation}
	D^0 \Phi^{(q)} = q \Phi^{(q)}\,.
\end{equation}

By construction, the harmonic variables are invariant under the AdS$^{4|8}$ isometry, 
\begin{equation}
\delta_\xi u^\pm_i = 0\,.
\label{delta-u}
\end{equation}
This means that they are inert under the $SU(2)$ generator $J_{ij}$. As a consequence, the $U(1)$ generator $ S^{ij}J_{ij}$ acting on the harmonic superfields (\ref{Phi-q}) is represented by the following operator:
\begin{align}
  {\cal J}\Phi^{(q)} := &S^{ij}J_{ij} \Phi^{(q)} 
  \nonumber\\=& (S^{--}D^{++} - S^{++}D^{--} + S^{+-}D^0)\Phi^{(q)}\,,
\end{align}
where 
\begin{equation}
S^{\pm\pm} = u^\pm_i u^\pm_j S^{ij}\,.
\end{equation}

Eq.~(\ref{delta-H}) implies the transformation law of the harmonic superfield $\Phi^{(q)}$ with no Lorentz indices under the isometry of AdS$^{4|8}$:
\begin{align}
\delta_\xi \Phi^{(q)} = &(\xi^a {\cal D}_a - \xi^{+\alpha}{\cal D^-_\alpha} + \xi^{-\alpha}{\cal D}^+_\alpha \nonumber\\&- \bar\xi^-_{\dot\alpha}\bar{\cal D}^{+\dot\alpha} + \bar\xi^+_{\dot\alpha}\bar{\cal D}^{-\dot\alpha} +\rho{\cal J})\Phi^{(q)}\,,
\label{delta-H-harmonic}
\end{align}
where
$\xi^{\pm\alpha} = u^\pm_i \xi^{\alpha i}$, $\bar\xi^\pm_{\dot\alpha} = u^\pm_i \bar\xi^i_{\dot\alpha}$, ${\cal D}^{\pm}_\alpha = u^\pm_i {\cal D}^i_\alpha$ and $\bar{\cal D}^\pm_{\dot\alpha} = u^\pm_i \bar{\cal D}^i_{\dot\alpha}$. 

For \emph{analytic} superfields obeying 
\begin{equation}
{\cal D}^+_\alpha \Phi_A^{(q)} = \bar{\cal D}^+_{\dot\alpha}\Phi_A^{(q)} = 0\,,
\label{analyticity}
\end{equation}
the variation (\ref{delta-H-harmonic}) simplifies to
\begin{equation}
\delta_\xi \Phi_A^{(q)} = (\xi^a {\cal D}_a - \xi^{+\alpha}{\cal D^-_\alpha} + \bar\xi^+_{\dot\alpha}\bar{\cal D}^{-\dot\alpha} + \rho{\cal J})\Phi_A^{(q)}\,.
\end{equation}
Note that the analyticity constraint (\ref{analyticity}) is preserved by the isometry transformations in AdS$^{4|8}$, ${\cal D}^+_\alpha \delta_\xi \Phi_A^{(q)} = \bar{\cal  D}^+_{\dot\alpha}\delta_\xi \Phi_A^{(q)} = 0$. This property follows from the definition of the Killing vector (\ref{KillingVectorEquation}) and the condition (\ref{delta-u}).

The algebra of covariant derivatives (\ref{D-algebra}) may be re-written in terms of covariant derivatives ${\cal D}^\pm_\alpha$ and $\bar{\cal D}^\pm_{\dot\alpha}$. Among these (anti)commutation relations we will need the following ones:
\begin{subequations}
\label{D+-algebra}
\begin{align}
\{ {\cal D}^+_\alpha, \bar{\cal D}^-_{\dot\alpha} \} = &-2i\sigma^a_{\alpha\dot\alpha} {\cal D}_a\,,\\
\{ {\cal D}^+_\alpha, {\cal D}^+_\beta \} = &4S^{++}M_{\alpha\beta}\,,\label{D+D+}\\
\{ \bar{\cal D}^+_{\dot\alpha} ,\bar{\cal D}^+_{\dot\beta} \} = &-4 S^{++}\bar M_{\dot\alpha\dot\beta}\,,\label{barD+D+}\\
\{ {\cal D}^+_{\alpha} , {\cal D}^-_\beta \} =& 4S^{+-}M_{\alpha\beta} -2\varepsilon_{\alpha\beta}{\cal J}\,, \\
\{ \bar{\cal D}^+_{\dot\alpha} , \bar{\cal D}^-_{\dot\beta} \} =& -4S^{+-}\bar M_{\dot\alpha\dot\beta} +2 \varepsilon_{\dot\alpha\dot\beta}{\cal J}\,,\\
[{\cal D}_a ,{\cal D}^\pm_\alpha] =& \frac i2(\sigma_a)_{\alpha\dot\alpha}(S^{\pm+}\bar{\cal D}^{-\dot\alpha} - S^{\pm-}\bar{\cal D}^{+\dot\alpha})\,,\\ 
[{\cal D}_a,\bar{\cal D}^\pm_{\dot\alpha}] = & \frac i2(\sigma_a)_{\alpha\dot\alpha}(S^{\pm+}{\cal D}^{-\alpha} - S^{\pm-}{\cal D}^{+\alpha})\,.
\end{align}
\end{subequations}

The anticommutators (\ref{D+D+},\ref{barD+D+}) imply the following identities for a general unconstrained harmonic superfield of the form (\ref{Phi-q})
\begin{align}
  {\cal D}^+_\alpha[({\cal D}^+)^2 + 4S^{++} ] \Phi^{(q)} =&0\,,\\
  \bar{\cal D}^+_{\dot\alpha}[(\bar{\cal D}^+)^2 + 4S^{++} ] \Phi^{(q)} =&0\,.
\end{align}
Hence the operator
\begin{equation}
\Pi^{(+4)} = \frac1{16}[({\cal D}^+)^2 + 4S^{++} ][(\bar{\cal D}^+)^2 + 4S^{++} ]
\label{AnalyticProjector}
\end{equation}
maps any full-superspace superfield to an analytic one.

As a corollary, the operator
\begin{equation}
\hat\Box:= \frac1{32}[({\cal D}^+)^2 + 4S^{++} ][(\bar{\cal D}^+)^2 + 4S^{++} ]D^{--}D^{--}
\end{equation}
preserves the analyticity and maps any analytic superfield to another analytic superfield. Using the algebra of covariant derivatives (\ref{D+-algebra}) it is possible to derive the following representation for this operator when it acts on an analytic superfield with $U(1)$ charge $q$:
\begin{align}
  \hat\Box\Phi_A^{(q)} =& \Big[
    {\cal D}^a{\cal D}_a - \frac14 S^{++}({\cal D}^-)^2 - \frac14 S^{++}(\bar{\cal D}^-)^2\nonumber\\& +\frac12 S^{++}S^{++}D^{--}D^{--}  \nonumber\\& 
    + {\cal J} ({\cal J} + S^{++}D^{--} - 3 S^{+-})    
    \Big]\Phi_A^{(q)}\,.
\end{align}
This operator is an AdS$_4$ counterpart of the flat $\cN=2$ harmonic superspace d'Alembert operator introduced in Ref.~\cite{Galperin:1985bj}.

\section{Analytic action principle}
\label{SecAction}

Given a real superfield Lagrangian ${\cal L} = {\cal L}(x,\theta,u)$ in the full harmonic superspace, it is possible to write the invariant action in the standard form
\begin{equation}
  S = \int d^{4|8}z \,E\int du\, {\cal L}\,,
  \label{Sfull}
\end{equation}
where $E = \mathrm{Ber}^{-1} E_A{}^M $ is the Berezinian of the inverse supervielbein. The non-trivial problem is, however, to construct invariant action in the \emph{analytic} subspace. 

Using the fact that the superspace integral of the full derivative is vanishing,
\begin{equation}
\int d^{4|8}z\,E\, {\cal D}^i_\alpha \Psi_i^\alpha = \int d^{4|8}z\,E\, \bar{\cal D}_i^{\dot\alpha}  \bar\Psi^i_{\dot\alpha} = 0\,,
\end{equation}
the action (\ref{Sfull}) may be expressed via the analytic projector (\ref{AnalyticProjector}):
\begin{equation}
S = \int d^{4|8} z\, E\int du \frac{{\cal L}^{(+4)}}{(S^{++})^2}\,,
\label{S4.3}
\end{equation}
where 
\begin{align}
{\cal L}^{(+4)} = & \frac1{16}[({\cal D}^+)^2 + 4S^{++} ][(\bar{\cal D}^+)^2 + 4S^{++} ] {\cal L} \nonumber\\ =& \Pi^{(+4)}{\cal L}
\end{align}
is an analytic Lagrangian. It obeys the reality condition under the tilde-conjugation in the harmonic superspace \cite{HSS}, $\tilde{\cal L}^{(+4)} = {\cal L}^{(+4)}$.

The analytic action principle in the form (\ref{S4.3}) was derived in the general setup of $\cN=2$ supergravity matter theory in the harmonic superspace in Ref.~\cite{Butter:2015nza}. It was proved there that this action is free from harmonic singularities for arbitrary covariantly analytic Lagrangian ${\cal L}^{(+4)}$ despite the fact that it contains the charged harmonic superfield $(S^{++})^2$ in the denominator. Here we apply the same action principle for matter theories on the particular AdS$^{4|8}$ background with the covariantly constant torsion superfield $S^{ij}$. 

More generally, the action principle for an analytic Lagrangian ${\cal L}^{(+4)}$ may be written as follows:
\begin{equation}
S = \int d^{4|8}z \,E\int du \frac{U}{\Pi^{(+4)}U} {\cal L}^{(+4)}\,,
\label{S4.4}
\end{equation}
where $U$ is an arbitrary nowhere vanishing superfield and $\Pi^{(+4)}U$ is its analytic projection. Integrating the covariant derivatives by parts, one can show that this action is independent of the particular choice of $U$, 
\begin{equation}
\delta_U S = 0\,.    
\end{equation}
Thus the superfield $U$ is pure gauge and the action (\ref{S4.4}) reduces to (\ref{S4.3}) for $U=1$. In Ref.~\cite{Butter:2015nza}, it was shown that this superfield $U$ originates from the vector multiplet compensator in the $\cN=2$ supergravity. See also Ref.~\cite{Review} for a review of covariant superspace approaches to $\cN = 2$ supergravity.

The action (\ref{S4.3}) is a harmonic superspace counterpart of the analogous action in the AdS$_4$ projective superspace proposed in Refs.~\cite{Kuzenko:2008qw,Kuzenko:2008ry} for an analytic Lagrangian ${\cal L}^{++}$:
\begin{equation}
    S = \frac1{2\pi}\oint(u^+ du^+)\int d^{4|8}z\, E\frac{{\cal L}^{++}}{(S^{++})^2}\,.
\end{equation}
Unlike the projective superspace approach, in the harmonic superspace one, the factor $(S^{++})^{-2}$ in the action (\ref{S4.3}) may represent a potential problem because of harmonic singularities. It is a non-trivial exercise to check that this factor cancels out upon calculation of the component structure of this action. As we will show below, there is an alternative action principle for an analytic Lagrangian where this problem does not occur.

Rather than further unfolding the action functional (\ref{S4.3}) we switch to a bottom-up approach: starting from the standard action principle in the flat harmonic superspace for an analytic Lagrangian \cite{HSS}, we seek for its direct AdS$_4$ harmonic superspace generalization. Indeed, given an analytic Lagrangian $L^{(+4)}$ in the flat harmonic superspace, the corresponding action may be written as $S=\int d^4x \int du(D^-)^4 L^{(+4)}|$, with flat superspace derivatives $D^-_\alpha, \bar D^-_{\dot\alpha}$ and `$|$' standing for the $\theta=0$ projection. In the curved superspace, the flat superspace derivatives should be replaced with the covariant ones, ${\cal D}^-_\alpha$ and $\bar{\cal D}^-_{\dot\alpha}$, and the Lagrangian is represented by a covariantly analytic superfield ${\cal L}^{(+4)}$. Then, considering the mass dimension of torsion superfield $S^{--}$ and balance of the $U(1)$ charge, it is natural to search for an AdS$_{4}$ generalization of the flat superspace action within the following most general ansatz:
\begin{subequations}
\label{S-action}
\begin{align}
S = & S_0 + a_1 S_1 + a_2 S_2\,,\label{S0+S1}\\
S_0 =& \int d^4x \,e \int du ({\cal D}^-)^4{\cal L}^{(+4)}| \,,\\
S_1 =& \int d^4x \,e \int du \, S^{--}[({\cal D}^-)^2 + (\bar{\cal D}^-)^2]{\cal L}^{(+4)}\big|\,,\\
S_2 =& \int d^4x \,e \int du \, (S^{--})^2{\cal L}^{(+4)} |,
\end{align}
\end{subequations}
where $\big|$ stands for the projection $\theta=0$, $d^4x\, e =d^4x \sqrt{-g}$ is purely bosonic AdS$_4$ invariant measure and
\begin{equation}
({\cal D}^-)^4 := \frac1{16}({\cal D}^-)^2 (\bar{\cal D}^-)^2\,.
\end{equation}
The coefficients $a_1$ and $a_2$ in (\ref{S0+S1}) are to be found from the condition of invariance of this action under the isometry of AdS$^{4|8}$, $\delta_\xi S =0$. 

Since the Lagrangian is a scalar, it transforms under (\ref{delta-H}) by the rule
\begin{equation}
  \delta_\xi {\cal L}^{(+4)} = (\xi + \rho {\cal J}){\cal L}^{(+4)}\,.
\end{equation}
Using the following identities for the integration by parts for a superfield $\Phi$
\begin{subequations}
\label{int-by-parts}
\begin{align}
  \int d^4x\,e \int du\,{\cal D}_a \Phi^{a} \big| = &0\,,\\
  \int d^4x\,e \int du\, D^{++}\Phi^{--}\big| =&0\,,\\
  \int d^4x\,e \int du\, D^{--}\Phi^{++}\big| =&0\,,\\
  \int d^4x\,e \int du\, {\cal J}\Phi^{(0)}\big| =&0\,,
\end{align}
\end{subequations}
we can represent the variations of $S_0$, $S_1$ and $S_2$ in the forms
\begin{subequations}
\label{deltaS0S1}
\begin{align}
  \delta_\xi S_0 =& \int d^4x\,e \int du( - \xi^{+\alpha}{\cal D^-_\alpha} + \xi^{-\alpha}{\cal D}^+_\alpha \nonumber\\& - \bar\xi^-_{\dot\alpha}\bar{\cal D}^{+\dot\alpha} + \bar\xi^+_{\dot\alpha}\bar{\cal D}^{-\dot\alpha} )({\cal D}^-)^4 {\cal L}^{(+4)} \big|. \\
  \delta_\xi S_1 =& \int d^4x\,e \int du\,S^{--} (
  - \xi^{+\alpha}{\cal D^-_\alpha} + \xi^{-\alpha}{\cal D}^+_\alpha - \bar\xi^-_{\dot\alpha}\bar{\cal D}^{+\dot\alpha} \nonumber\\&  + \bar\xi^+_{\dot\alpha}\bar{\cal D}^{-\dot\alpha} 
  )[({\cal D}^-)^2 + (\bar{\cal D}^-)^2] {\cal L}^{(+4)}\big|\,,\\
  \delta_\xi S_2 =& \int d^4x\,e \int du(S^{--})^2(\bar\xi^+_{\dot\alpha}\bar{\cal D}^{-\dot\alpha} - \xi^{+\alpha}{\cal D}^-_\alpha){\cal L}^{(+4)}\big|\,.
\end{align}
\end{subequations}
We will proceed by the following steps:
\begin{enumerate}
\item Commute the derivatives ${\cal D}^+_\alpha$ and $\bar{\cal D}^{+\dot\alpha}$ with $({\cal D}^-)^2$ and $(\bar{\cal D}^-)^2$ using the algebra of covariant derivatives (\ref{D+-algebra}) and the analyticity of the Lagrangian, ${\cal D}^+_\alpha{\cal L}^{(+4)} = \bar{\cal D}^{+\dot\alpha}{\cal L}^{(+4)} =0$.
\item Using the same algebra of covariant derivatives, in the integrand pull the operators ${\cal D}_a$ and $\cal J$ to the left and use the rules of integration by parts (\ref{int-by-parts}). 
\item When the derivative ${\cal D}_a$ hits the Killing spinors, the following identities may be applied:
\begin{align}
  \sigma^a_{\alpha\dot\alpha} {\cal D}_a \xi^{-\alpha} =&2i (S^{--}\bar\xi^+_{\dot\alpha} - S^{+-}\bar\xi^-_{\dot\alpha})\,,\\
  (\sigma^a)^{\alpha\dot\alpha} {\cal D}_a \bar\xi^-_{\dot\alpha}=&-2i(S^{--}\xi^{+\alpha} - S^{+-}\xi^{-\alpha})\,.
\end{align}
These identities follow from Eq.~(\ref{xi-equation}).
\end{enumerate}
As a result, we find the following expressions for the variations (\ref{deltaS0S1}):
\begin{align}
\delta_\xi S_0 =& \frac12\int d^4x \,e\int du\,S^{--}
 \Big[ \xi^{+\alpha}{\cal D}^-_\alpha (\bar{\cal D}^-)^2
 \nonumber\\& - \bar\xi^+_{\dot\alpha}\bar{\cal D}^{-\dot\alpha}({\cal D}^-)^2 - 8S^{--}\xi^{+\alpha}{\cal D}^-_\alpha + 8S^{--}\bar\xi^+_{\dot\alpha}\bar{\cal D}^{-\dot\alpha}
 \nonumber \\&
 +12S^{+-}\xi^{-\alpha}{\cal D}^-_\alpha - 12S^{+-}\bar\xi^-_{\dot\alpha}\bar{\cal D}^{-\dot\alpha}
 \Big]{\cal L}^{(+4)}\big|\,,\\
 \delta_\xi S_1 =& -\int d^4x \,e\int du\,S^{--}
 \Big[ \xi^{+\alpha}{\cal D}^-_\alpha (\bar{\cal D}^-)^2 \nonumber\\&- \bar\xi^+_{\dot\alpha}\bar{\cal D}^{-\dot\alpha}({\cal D}^-)^2 - 8S^{--}\xi^{+\alpha}{\cal D}^-_\alpha + 8S^{--}\bar\xi^+_{\dot\alpha}\bar{\cal D}^{-\dot\alpha}
 \nonumber \\&
 +12S^{+-}\xi^{-\alpha}{\cal D}^-_\alpha - 12S^{+-}\bar\xi^-_{\dot\alpha}\bar{\cal D}^{-\dot\alpha}
 \Big]{\cal L}^{(+4)}\big|\,.
\end{align}
Hence, for $a_1=1/2$ and $a_2 = 0$, the action (\ref{S-action}) is invariant under the isometry of AdS$^{4|8}$, $\delta_\xi(S_0+\frac12 S_1)=0$. Finally, we write the action principle for an analytic Lagrangian ${\cal L}^{(+4)}$ in the form:
\begin{align}
S = &\int d^4x \,e \int du \bigg\{ ({\cal D}^-)^4
\nonumber\\&
+\frac12 S^{--} \left[({\cal D}^-)^2 + (\bar{\cal D}^-)^2\right]
\bigg\}{\cal L}^{(+4)}\Big|\,.
\label{Saction}
\end{align}
In this form, the analytic action principle is free from harmonic singularities and is suitable for studying the component structure of the analytic Lagrangian ${\cal L}^{(+4)}$. The relation of this action to Eq.~(\ref{S4.3}) will be studied elsewhere.

Below, we demonstrate how the analytic action principle (\ref{Saction}) applies for calculation of the component structure in the free $q$-hypermultiplet model.


\section{Example: component structure of the free hypermultiplet action}

The free $q$-hypermultiplet model is described by the following Lagrangian
\begin{equation}
    {\cal L}^{(+4)} = -\tilde q^+ D^{++}q^+\,,
    \label{Lq}
\end{equation}
where $q^+$ is an analytic superfield and $\tilde q^+$ is its tilde-conjugate,
\begin{equation}
    {\cal D}^+_\alpha q^+ = \bar{\cal D}^+_{\dot\alpha} q^+ = 0\,.\label{q-analytic}
\end{equation}
Note that this superfield is not required to be annihilated by $D^{++}$ and, thus, is, in general, represented by an infinite series over the harmonic variables. This feature is crucial for the off-shell description of the $q$-hypermultiplet model in the harmonic superspace.

To evaluate the component structure of the action (\ref{Saction}) with the Lagrangian (\ref{Lq}), we proceed by the following steps:
\begin{enumerate}
		\item We commute the derivatives $\mathcal{D}^-_\alpha$ and $\bar{\mathcal{D}}^{-\dot\alpha}$ to the right and act on $q^+$ and $\tilde q^+$. The commutators 
        $[{\cal D}^-_\alpha,D^{++}] = -{\cal D}^+_\alpha$ and $[\bar{\cal D}^{-\dot\alpha},D^{++}] = - \bar{\cal D}^{+\dot\alpha}$
        yields the derivatives $\mathcal{D}^+$'s, which annihilate the analytic superfield $q^+$ according to Eq.~(\ref{q-analytic}).
		\item The operators ${\cal D}_a$ and $\cal J$ are moved to the left by using the rules for integration by parts (\ref{int-by-parts}).
	\end{enumerate}
As a result, we obtain the following bosonic and fermionic field contributions to the free hypermultiplet action
\begin{widetext}
\begin{align}
    S_q= &S_\mathrm{bos}+S_\mathrm{ferm}\,,\\
	S_\mathrm{bos} =& -\int d^4x \,e \int du \Bigl[\bigl((\mathcal{D}^-)^4\tilde{q}^+\bigr) D^{++} q^+ +\tilde{q}^+ D^{++} \bigl((\mathcal{D}^-)^4 q^+\bigr)  + \frac{1}{4} (\mathcal{D}^{-\beta}\bar{\mathcal{D}}^{-\dot{\alpha}} \tilde{q}^+) D^{++} (\mathcal{D}^-_\beta \bar{\mathcal{D}}^-_{\dot\alpha} q^+) \nonumber\\ & +\frac{1}{16} \bigl((\bar{\mathcal{D}}^-)^2\tilde{q}^+\bigr) D^{++} ((\mathcal{D}^-)^2 q^+) +\frac{1}{16} \bigl((\mathcal{D}^-)^2\tilde{q}^+\bigr) D^{++} (({\bar{\mathcal{D}}}^-)^2 q^+) \nonumber\\& +\frac{i}{2} (\mathcal{D}^{-\beta}\bar{\mathcal{D}}^{-\dot\alpha} \tilde{q}^+) (\mathcal{D}_{\beta\dot\alpha} q^+) - \frac{i}{2} (\mathcal{D}_{\beta\dot\alpha}\tilde{q}^+) (\mathcal{D}^{-\beta}\bar{\mathcal{D}}^{-\dot\alpha} q^+) +\frac{1}{4} \bigl((\mathcal{D}^-)^2\tilde{q}^+ +(\bar{\mathcal{D}}^-)^2\tilde{q}^+\bigr) (\mathcal{J} + 2 S^{--} D^{++}) q^+ \nonumber\\& +\frac{1}{4} \tilde{q}^+ (\mathcal{J} +2 D^{++}S^{--})\bigl((\mathcal{D}^-)^2 q^+ +(\bar{\mathcal{D}}^-)^2 q^+\bigr) + 9 S^{--} \tilde{q}^+ \mathcal{J}q^+\Bigr]\bigg|\,,
    \label{Sbos}\\
%
S_\mathrm{ferm} =& -\int d^4x \,e \int du 
	\Bigl[\frac{1}{8} (\mathcal{D}^{-\alpha}(\bar{\mathcal{D}}^-)^2\tilde{q}^+) D^{++} (\mathcal{D}^-_\alpha q^+) +\frac{1}{8} (\bar{\mathcal{D}}^-_{\dot\beta}\tilde{q}^+) D^{++} (\bar{\mathcal{D}}^{-\dot\beta}(\mathcal{D}^-)^2 q^+)\nonumber\\ & +\frac{1}{8} (\bar{\mathcal{D}}^-_{\dot\beta}(\mathcal{D}^-)^2\tilde{q}^+) D^{++} (\bar{\mathcal{D}}^{-\dot\beta} q^+) 
    + \frac{1}{8} (\mathcal{D}^{-\alpha} \tilde{q}^+) D^{++} (\mathcal{D}^-_\alpha (\bar{\mathcal{D}}^-)^2 q^+) \nonumber\\& +\frac{i}{2} (\mathcal{D}^{-\alpha} \tilde{q}^+) (\mathcal{D}_{\alpha\dot\beta}\bar{\mathcal{D}}^{-\dot\beta}q^+) 
    - \frac{i}{2} (\bar{\mathcal{D}}^{-\dot\beta} \tilde{q}^+) (\mathcal{D}_{\alpha\dot\beta}\mathcal{D}^{-\alpha} q^+)  
    + \frac{1}{2} (\mathcal{D}^{-\alpha}\tilde{q}^+) (\mathcal{J}+ 2 S^{+-} +2 S^{--} D^{++}) (\mathcal{D}^-_\alpha q^+)\nonumber\\& +\frac{1}{2} (\bar{\mathcal{D}}^-_{\dot\beta}\tilde{q}^+) (\mathcal{J}+ 2 S^{+-} +2 S^{--} D^{++}) (\bar{\mathcal{D}}^{-\dot\beta} q^+)\Bigr]\bigg|\,.
    \label{Sferm}
\end{align}
Next, we define the bosonic and fermionic components of the hypermultiplet superfield with the following equations:
\begin{subequations}
\begin{align}
    q^+| = & F^+(x,u)\,,& \tilde q^+| = & -\bar F^+(x,u)\,,\\
    {\cal D}^-_\alpha q^+| =& \Psi_\alpha(x,u)\,,& \bar{\cal D}^{-\dot\alpha}\tilde q^+| = & -\bar\Psi^{\dot\alpha}(x,u)\,,\\
    \bar{\cal D}^{-\dot\alpha}q^+| =&\bar\Phi^{\dot\alpha}(x,u)\,,& {\cal D}^-_\alpha \tilde q^+| =& \Phi_\alpha(x,u)\,,\\
    ({\cal D}^-)^2 q^+| = &-4M^-(x,u)\,,& (\bar{\cal D}^-)^2 \tilde q^+| =& 4\bar M^-(x,u)\,,\\
    (\bar{\cal D}^-)^2 q^+| =&-4N^-(x,u)\,, & ({\cal D}^-)^2 \tilde q^+| =& 4\bar N^-(x,u)\,,\\
    {\cal D}^-_\beta \bar{\cal D}^-_{\dot\alpha} q^+| =& -i\sigma^a_{\beta\dot\alpha} A^-_a(x,u) = -iA_{\beta\dot\alpha}^-(x,u)\,,&
    {\cal D}^-_\beta\bar{\cal D}^-_{\dot\alpha}\tilde q^+| =& i\sigma^a_{\beta\dot\alpha }\bar A^-_a(x,u) = i\bar A^-_{\beta\dot\alpha}(x,u)\,,\\
    {\cal D}^-_\alpha(\bar{\cal D}^-)^2 q^+| =& -4\Lambda_\alpha^{--}(x,u)\,, & \bar{\cal D}^{-\dot\alpha}({\cal D}^-)^2 \tilde q^+| =& 4\bar\Lambda^{--\dot\alpha}(x,u)\,,\\
    \bar{\cal D}^{-\dot\alpha}({\cal D}^-)^2 q^+| =&-4\bar X^{--\dot\alpha}(x,u)\,,& {\cal D}^-_\alpha (\bar{\cal D}^-)^2 \tilde q^+| =&-4X_\alpha^{--}(x,u)\,,\\
    ({\cal D}^-)^4 q^+|=&P^{(-3)}(x,u)\,, & ({\cal D}^-)^4\tilde q^+| =&-\bar P^{(-3)}(x,u)\,.
\end{align}
\end{subequations}
These components enter the actions (\ref{Sbos}) and (\ref{Sferm}) as follows:
\begin{subequations}
\begin{align}
    S_\mathrm{bos} =& \int d^4x \,e\int du\Big[
     \bar P^{(-3)}D^{++}F^+ + \bar F^+ D^{++} P^{(-3)} + \frac12 \bar A^-_a D^{++} A^{-a} + 9S^{--} \bar F^+ {\cal J}F^+ \nonumber\\ 
     &+\bar M^- D^{++}M^- + \bar N^- D^{++} N^- -\bar A_a({\cal D}^a F^+) + ({\cal D}_a \bar F^+)A^{-a} \nonumber\\
     &-(\bar M^- + \bar N^-)({\cal J} + 2 S^{--}D^{++})F^+
     -\bar F^+({\cal J} +2 D^{++}S^{--})(M^- + N^-)
    \Big],\\
    S_\mathrm{ferm} =& \int d^4x\,e\int du\Big[
    -\frac i2 \Phi^\alpha ({\cal D}_{\alpha\dot\beta}\bar\Phi^{\dot\beta}) - \frac i2 \bar\Psi^{\dot\beta}({\cal D}_{\alpha\dot\beta}\Psi^\alpha)\nonumber\\
    &+\frac12 X^{--\alpha}D^{++}\Psi_\alpha - \frac12\bar\Psi_{\dot\alpha} D^{++}\bar X^{--\dot\alpha} - \frac12\bar\Lambda^{--}_{\dot\alpha}D^{++}\bar\Phi^{\dot\alpha} + \frac12\Phi^\alpha D^{++}\Lambda^{--}_\alpha\nonumber\\
    &-\frac12\Phi^\alpha({\cal J} + 2S^{+-} + 2S^{--}D^{++})\Psi_\alpha
    +\frac12\bar\Psi_{\dot\alpha}({\cal J} + 2S^{+-} + 2S^{--}D^{++})\bar\Phi^{\dot\alpha}
    \Big].
\end{align}
\end{subequations}
The corresponding equations of motion for these component fields are:
\begin{subequations}
\begin{align}
    D^{++}F^+ =&0\,,\label{eq-a}\\
    D^{++}M^- - ({\cal J} + 2S^{--}D^{++})F^+ =&0\,,\label{eq-b}\\
    D^{++}N^- - ({\cal J} + 2S^{--}D^{++})F^+ =&0\,,\label{eq-c}\\
    D^{++}A^-_a - 2{\cal D}_a F^+ =&0\,,\label{eq-d}\\
    D^{++}\Psi_\alpha =0\,,\qquad
    D^{++}\Phi_\alpha =&0\,,\label{eq-e}\\
    D^{++}P^{(-3)} - {\cal D}^a A^-_a + 9S^{--}{\cal J}F^+ - ({\cal J} + 2D^{++}S^{--})(M^- + N^-) =&0\,,\label{eq-f}\\
    D^{++}\Lambda^{--}_\alpha - i{\cal D}_{\alpha\dot\beta}\bar\Phi^{\dot\beta} - ({\cal J} + 2S^{+-} + 2S^{--}D^{++})\Psi_\alpha =&0\,,\\
    D^{++}X^{--}_\alpha + i{\cal D}_{\alpha\dot\beta} \bar\Psi^{\dot\beta} - ({\cal J} + 2S^{+-} + 2S^{--}D^{++})\Phi_\alpha =&0\,.\label{eq-h}
\end{align}
\end{subequations}
\end{widetext}
Elimination of auxiliary fields from these equations follows the standard procedure \cite{HSS}: equations (\ref{eq-a})--(\ref{eq-e}) are purely kinematic and so may be solved by expanding the series in harmonic variables and comparing term by term, thus eliminating most of the auxiliary fields. Equations (\ref{eq-f})--(\ref{eq-h}) represent a combination of kinematic and dynamic equations and allow one to eliminate the rest of the auxiliary fields, and are also responsible for putting the physical fields on shell. For instance, Eq.~(\ref{eq-a}) implies $F^+(x,u) = f^i u^+_i$; Eqs.~(\ref{eq-b}--\ref{eq-d}) imply $M^-(x,u) = M^i(x) u^-_i$, $N^-(x,u) = N^i(x) u^-_i$, $A^-_a(x,u) = A^i_a(x) u^-_i$, e.t.c. Note that the fermionic fields $\Psi_\alpha$ and $\Phi_\alpha$ are harmonic independent, $\Psi_\alpha(x,u) = \psi_\alpha(x)$, $\Phi_\alpha(x,u) = \phi_\alpha(x)$. Eliminating all remaining auxiliary fields via their equations of motion and evaluating integrals over harmonic variables, we find the following actions for the physical components:
\begin{align}
    S_q = &\int d^4x \,e\Big[ {\cal D}_m \bar f_i {\cal D}^m f^i - 2S^2 \bar f_i f^i
    \nonumber\\& +\frac i2 \bar\phi^{\dot\beta}{\cal D}_{\alpha\dot\beta}\phi^\alpha + \frac i2 \bar\psi^{\dot\beta}{\cal D}_{\alpha\dot\beta}\psi^\alpha
    \Big],
    \label{Sbos-result}
\end{align}

According to Eq.~(\ref{curvature}), the Riemann curvature tensor in the AdS$_4$ space is $R_{ab}{}^{cd} = -2S^2 \delta^c_{[a}\delta^d_{b]}$, and, hence, the Ricci scalar is $R = -12 S^2$. This allows us to match the AdS$_4$ mass term for the scalars in Eq.~(\ref{Sbos-result}) with the one in Ref.~\cite{Ivanov:2025jdp}, $-2S^2 = R/6$.

In conclusion, it would be interesting to study whether the superfield action (\ref{Saction}) can be obtained by dimensional reduction from the superfield theory on AdS$^{5|8}$ developed in Refs.~\cite{Kuzenko:2007aj,Kuzenko:2007vs}. In curved space, dimensional reduction is a non-trivial problem and certainly merits future investigation.

\vspace{5mm}
{\em Acknowledgments.}
We are grateful to Sergei Kuzenko for proposing to study the problem of analytic action principle in the AdS$_4$ harmonic superspace and for useful comments. IBS acknowledges the support from the Australian Research Council, project No.~DP230101629 during his work at UWA.


\end{document}